\documentclass[prb,twocolumn,showpacs]{revtex4}
\usepackage{graphicx}
\usepackage{amsmath}
\usepackage{amssymb}
\usepackage{dcolumn}
\usepackage{float}
\usepackage{bm}

\usepackage[utf8]{inputenc}
\usepackage[T1]{fontenc}

\newcommand{\e}{\mathrm{e}}

\DeclareMathAlphabet{\bi}{OML}{cmm}{b}{it}

\def\be{\begin{equation}}
\def\ee{\end{equation}}
\def\bearr{\begin{eqnarray}}
\def\eearr{\end{eqnarray}}
\def\la{\langle}
\def\ra{\rangle}

\begin{document}
%\title{Electron-phonon scattering in a 2DEG with spin-orbit interaction}
\title{Acoustic phonon-limited resistivity in spin-orbit coupled 2DEG:
Deformation potential and piezoelectric scattering}
\bigskip
\author{Tutul Biswas and Tarun Kanti Ghosh}
\normalsize
\affiliation
{Department of Physics, Indian Institute of Technology-Kanpur,
Kanpur-208 016, India}
\date{\today}

\begin{abstract}
We study the interaction between electron and acoustic phonon in a
Rashba spin-orbit coupled two dimensional electron gas using Boltzmann 
transport theory. Both deformation 
potential and piezoelectric scattering mechanisms are considered in 
the Bloch-Gruneisen (BG) as well as in the equipartition (EP) regimes. 
Effect of the Rashba spin-orbit interaction on the temperature dependence 
of resistivity in the BG and EP regimes has been discussed.
We find effective exponent of the temperature dependence 
of the resistivity in the BG regime decreases due to spin-orbit coupling.

\end{abstract}
\pacs{71.38.-k, 71.70.Ej, 72.20.Dp}

%71.38.-k: Polarons and electron-phonon interactions 
%(see also 63.20.K- Phonon interactions in lattice dynamics)
%71.70.Ej:Spin-orbit coupling, Zeeman and Stark splitting, Jahn-Teller effect 
%72.20.Dp:General theory, scattering mechanisms 
%72.25.Dc:Spin polarized transport in semiconductors
 
\maketitle
\section{Introduction}
With the rise of the promising field of spintronics, 
two dimensional electron gas (2DEG) with spin-orbit interaction (SOI) 
in semiconductor heterostructures has drawn much attention 
in recent years \cite{winkler,zutic,cahay}. The importance of this field was first 
realized when Datta and Das gave a proposal of Spin Field Effect 
Transistor \cite{datta}. Several studies on spintronics have been 
performed in recent years from both theoretical and experimental viewpoints. 
One major aim is to manipulate the spin degree of freedom of charge carriers 
in semiconductor nanostructures \cite{fabian} so that spin-based device 
technology \cite{wolf} and quantum information processing \cite{david} 
can be developed in near future. The SOI is an intrinsic phenomena
present in semiconductors. Mainly, there are two kinds of SOI present in
the semiconductor heterostructures we come across in the literature. 
One of them is the Rashba spin-orbit interaction 
(RSOI) \cite{rashba} which originates from the inversion asymmetry of the 
confining potential in semiconductor heterostructures. 
The RSOI is proportional to the magnitude of the electric field internally 
generated due to the band bending or externally applied gate voltage. 
It can be tuned by applying a gate voltage \cite{nitta,mats}.
Another kind of SOI is the Dresselhaus spin-orbit interaction \cite{dress} 
which originates from the bulk inversion asymmetry of the host crystal. 
It entirely depends on crystal property and it is not tunable.

Various electronic and transport properties of a 2DEG will be modified 
in the presence of the SOI. We mainly focus  on the RSOI in 2DEG systems
such as AlGaAs/GaAs heterostructure.
The coupling between electron and phonon
becomes stronger when the SO coupling constant is large enough and due
to this fact the effective mass is also increased \cite{cape}. The
critical temperature of  superconductors can be controlled by RSOI 
when the Fermi energy is small as compared to the
characteristic energy scale of RSOI \cite{grima}. 
The RSOI can produce infinite number of bound states 
in a 2DEG with short-range impurity potentials \cite{chap}. 
At low frequency
in the  presence of RSOI the universality of spin Hall conductivity
can be broken by considering the contribution of electron-phonon 
interaction to the spin-vertex correction \cite{mars}. The mobility, polaron mass 
correction and polaron binding energy can be changed significantly due to 
the presence of the RSOI \cite{chen,zhang}.
The relaxation time for various impurity potentials of low-dimensional semiconductor
structures with SOI has been studied \cite{pagano}.

The interaction between electron and phonon plays a very
crucial role in determining the transport properties of a 2DEG and it has a 
finite contribution to the momentum relaxation time of the charge 
carriers. Other contributions \cite{knight,jiang,gold,dolgo,loss} 
come from disorders, impurities, etc. 
There are three distinct temperature regimes: a) Bloch-
Gruneisen (BG), b) Equipartition (EP) and c) Inelastic phonon scattering.
The BG temperature $T_{BG}$ can be defined \cite{stormer} by the relation
$k_BT_{BG}=2\hbar k_Fv_s$, where $k_B$ is the Boltzmann constant, $v_s$ 
is the phonon velocity and $k_F$ is the Fermi wave vector. 
For typical electron density ($n_e \sim 10^{15}$ m${}^{-2}$) in a 2DEG system, 
$T_{BG}$ is around $6.2 $ K. 
In the BG regime, a direct manifestation of the acoustic 
phonon-dominated transport property is the strong change in the temperature
dependence of the resistivity.  
The existence of the BG regime has been confirmed for 2DEG
experimentally \cite{stormer}. 
The problem of electron-phonon interaction in a 2DEG confined in
semiconductor heterostructures has been studied extensively
\cite{price1,price2,price3,ridley,price4,price5,dassarma,mos2}.
Recently, the phonon-dominated transport properties in the BG regime 
have been studied in graphene both theoretically \cite{graphene_t,castro,kris} and 
experimentally \cite{graphene_exp}. 
%To the best of our knowledge, a detailed study of the effect of
%the RSOI coupling on the momentum relaxation time due to electron-phonon
%scattering has not been done yet.

In the present work we would like to investigate the influence of the RSOI
on momentum relaxation time due to electron-phonon interaction and 
hence on the transport properties of 2DEG systems. We consider both the cases for 
perfect 2DEG and quasi-2DEG (usually found in semiconductor heterostructures). 
In the former case 2D phonon wave vector ${\bf q}$ couples with 
2D electron wave vector ${\bf k}$ and in the latter case the coupling 
between 3D bulk phonon wave vector ${\bf Q} = ({\bf q}, q_z)$ and 2D electron 
wave vector
${\bf k}$ is considered. We consider two mechanisms of electron-phonon 
interaction, namely deformation potential (DP) and piezoelectric (PE) potential 
scattering separately. In all the cases linear temperature dependence of 
inverse relaxation time (IRT) is found in the high temperature (EP) regime.
But in the BG regime for perfect 2DEG and quasi-2DEG we find analytically
resistivity is proportional to $T^4$ and $T^5$, respectively, in the case of DP
scattering mechanism. 
On the other hand, $\rho \sim T^3$ for PE scattering mechanism in a quasi-2DEG. 
Our numerical calculations reveal that this exponent of $T$ 
{\it strongly} depends on density and the SO coupling constant. In fact,
the exponent of $T$ decreases due to presence of spin-orbit coupling. 
We also discuss the resistivity as a function of the SO coupling constant
in the BG and EP regimes.

This paper is organized as follows. In section II 
we derive all the theoretical results and discuss all the 
numerical analysis for perfect 2DEG. In section III both DP and PE 
scattering mechanisms have been taken into account for quasi-2DEG
and we discuss analytical and numerical results in detail.
We summarize all the results in section IV.

\section{Electron-phonon scattering in a perfect 2DEG}
\subsection{Theoretical model}
We consider a 2DEG with the RSOI in the $xy$ plane in a semiconductor 
heterostructure.
The single-particle Hamiltonian of this system is given by
\begin{eqnarray}
H=\frac{{\bf p}^2}{2m^\ast} \sigma_0  +
\frac{\alpha}{\hbar}\big({\sigma}_x{p}_y-{\sigma}_y{p}_x\big),
\end{eqnarray}
where ${\bf p} $  is the two-dimensional momentum operator, 
$ m^{\ast} $ is the effective mass of an electron, $ \sigma_0 $
is the unit $2 \times 2$ matrix,
$\alpha$ is the Rashba spin-orbit coupling constant and 
$ \sigma_{x(y)}$ are the Pauli matrices.
The eigenenergies are given by
\begin{equation} \label{energy}
\epsilon_{\lambda}(k)=\frac{\hbar^2k^2}{2m^\ast} + 
\lambda \alpha | {\bf k} |,
\end{equation}
with the corresponding normalized eigenspinors 
\begin{eqnarray} \label{spinstates}
\psi_{\lambda} (x,y) = \frac{1}{\sqrt{2A}}\begin{pmatrix} 1
\\ \lambda e^{-i \phi} \end{pmatrix} e^{i {\bf k} \cdot {\bf r}}. 
\end{eqnarray}
Here, $ \lambda = \pm $ represents the upper and lower energy branches,
$A$ is the area of the system
and $ \phi = \tan^{-1}(k_x/k_y)$.
The density of states for the two energy branches 
are given by \cite{cape}
\begin{eqnarray}
D_{\pm}(\epsilon)&=&\frac{D_0}{2}\Big(1 \mp
\sqrt{\frac{\epsilon_\alpha}{\epsilon+\epsilon_\alpha}}\Big)
\Theta(\epsilon)\nonumber\\
& + & D_0\sqrt{\frac{\epsilon_\alpha}{\epsilon+\epsilon_\alpha}}
\Theta(-\epsilon)\Theta(\epsilon+\epsilon_\alpha).
\end{eqnarray}
Here, $ D_0 = m^*/(\pi \hbar^2)$, $ \Theta(x) $ is the unit
step function and $\epsilon_{\alpha}=m^\ast\alpha^2/(2\hbar^2)$ is 
the characteristic energy scale of RSOI.  

The Hamiltonian for electron-phonon interaction in the case of 
deformation potential coupling can be written as 
$H_{\rm ep} = D {\bf{\nabla}} \cdot \bf{u({\bf r})}$,
where $ D$ is the deformation-potential coupling constant and
the lattice displacement vector {\bf u({\bf r})} is given by
\begin{equation}
{\bf u({\bf r})}=\sum_{\bf q} \sqrt{\frac{\hbar}{2MN\omega_{\bf q}}}{\bf e}_q
[a_{\bf q} e^{i{\bf q}\cdot{\bf r}}+a^{\dagger}_{\bf q} e^{-i{\bf q \cdot {\bf r}}}]. 
\end{equation}
Here, $ \omega_{\bf q} = v_s q $ is the phonon frequency with the wave vector $q$ and
the sound velocity $v_s$, $a^{\dagger}_{\bf q}$ and $a_{\bf q}$ are phonon creation 
and annihilation operators, respectively.
Also, $ {\bf e}_q $ is a unit vector in the direction of the phonon
polarization.
The corresponding Hamiltonian can be written as
\begin{equation} \label{pot}
 H_{\rm ep}({\bf r}) = \sum_{\bf q}
\big[C_{\bf q}a_{\bf q}e^{i{\bf q\cdot{\bf r}}} + 
C^{\dagger}_{\bf q} a^{\dagger}_{\bf q}e^{-i{\bf q}\cdot {\bf r}}\big],
\end{equation}
where $C_{\bf q} = D \sqrt{\hbar/2M N\omega_{\bf q}}(i{\bf e}_q\cdot {\bf q})$.

The energy dependent relaxation time for electrons in a given
energy branch $\lambda $ can be written as
\begin{equation} \label{relax}
\frac{1}{\tau^{\lambda} (\epsilon)} = \sum_{{\bf k^\prime},\lambda^{\prime}}
(1-\cos{\theta_{\bf kk^\prime}})P_{\bf kk^\prime}^{\lambda \lambda^{\prime}}
\frac{1-f(\epsilon_{\lambda^\prime}(k^{\prime}))}{1-f(\epsilon_{\lambda}(k))},
\end{equation}
where $\theta_{\bf kk^\prime} $ is the scattering angle between
the two momentum vectors ${\bf k}$ and $\bf k^\prime$, 
$\epsilon_{\lambda}$ is given by
Eq. (\ref{energy}), $P_{\bf kk^\prime}^{\lambda \lambda^{\prime}}$ 
is the transition rate for scattering of an electron 
from a state $ |\bf k, \lambda \ra$ to 
$ | \bf k^\prime, \lambda^{\prime} \ra $
and $ f(\epsilon) = [\e^{\beta(\epsilon-\mu)} + 1]^{-1}$ is the 
Fermi-Dirac distribution function with $\beta = 1/(k_{B}T)$.
The chemical potential $\mu$ at finite temperature $T$ can be obtained 
self-consistently from the following normalization condition \cite{gao}:
\begin{eqnarray}\label{chem}
n_e=\int_0^\infty d\epsilon D_0f(\epsilon)+
\int_{-\epsilon_\alpha}^0 d\epsilon D_0f(\epsilon)
\sqrt{\frac{\epsilon_\alpha}{\epsilon+\epsilon_\alpha}},
\end{eqnarray}
where $n_e$ is the electron density. 
At $T=0$, the above equation reduces to 
$ \epsilon_F = \epsilon_F^0 - 2 \epsilon_{\alpha} $, where
$\epsilon_F $ and $\epsilon_F^0 $ is the Fermi energy of a 2DEG 
in presence and absence of RSOI, respectively.
Thus, the reduction in the Fermi energy due to RSOI is 
$ 2\epsilon_{\alpha}$.

We consider the interaction between electron and acoustic 
phonon due to deformation potential coupling.
The transition rate due to electron-phonon interaction can be 
written as
\begin{eqnarray} \label{trans_rate}
P_{\bf kk^\prime}^{\lambda \lambda^{\prime}} & = & \frac{2\pi}{\hbar} \sum_{\bf q} 
\vert C_{\bf q}^{\lambda \lambda^{\prime}} \vert^2 
\Big[ N_{\bf q}\delta(\epsilon_{\lambda^{\prime}}^\prime -
\epsilon_\lambda - \hbar\omega_{\bf q}) \nonumber \\
& + & 
(N_{\bf q}+1)\delta(\epsilon_{\lambda^{\prime}}^\prime -
\epsilon_\lambda + \hbar\omega_{\bf q}) \Big],
\end{eqnarray}
where $C_{\bf q}^{\lambda \lambda^{\prime}}$ is the matrix element for
the acoustic phonon and is given by
\begin{equation}
\vert C_{\bf q}^{\lambda \lambda^{\prime}}
\vert^2=\frac{D^2\hbar q}{2A\rho_a v_s}
\frac{1 + \lambda \lambda^{\prime} \cos \theta}{2} 
\delta_{\lambda\lambda^{\prime}}.
\end{equation}
Here,  $ \theta \equiv  \theta_{\bf kk^{\prime}}$, 
$D$ is the deformation potential coupling constant
and $\rho_{a} = NM/A$ is the mass per unit area.
The appearance of $ \delta_{\lambda \lambda^{\prime}} $ is due to
the fact that the electron-phonon interaction given by Eq. (\ref{pot}) 
is spin-independent.
Also, $ N_{\bf q} = [\exp( \beta \hbar\omega_{\bf q} )-1]^{-1} $ is the
phonon occupation number.
The first and second terms on the right hand side of 
Eq. (\ref{trans_rate})  correspond to the absorption and emission 
of a phonon with energy $\hbar\omega_{\bf q}$, respectively.
Within the small-angle scattering approximation ($q = 2k \sin (\theta/2)$),
the matrix element for intra-branch scattering 
($ \lambda = \lambda^{\prime}$) becomes
$\vert C_{\bf q}^{\lambda \lambda^{\prime}}
\vert^2=\frac{D^2\hbar q}{2A \rho_{a} v_s}
(1- \frac{q^2}{4 k^2}) $. 
The similar matrix element is obtained for a single layer graphene
\cite{graphene_t}.
%We note that the matrix element due to DP coupling of a 2DEG without
%RSOI can not be obtained by taking $\alpha=0$.

Before presenting the numerical results, we present 
how IRT depends on $T$ in the EP and the BG regimes.
At high temperature (EP regime), the phonon energy is much smaller
than the thermal energy i.e.  $\hbar\omega_{\bf q}\ll k_{B}T$. 
We neglect $\hbar\omega_{\bf q}$ term in the delta functions 
i.e.  $\delta(\epsilon^\prime-\epsilon\mp\hbar\omega_{\bf q}) 
\simeq \delta(\epsilon^\prime-\epsilon)$.
Again, in this temperature limit the Bose occupation factor 
$N_{\bf q}$ can be
approximated as 
$ N_{\bf q} \simeq N_{\bf q} + 1 \simeq  k_{B}T/\hbar\omega_{\bf q}$.
When $ \epsilon_{F} \geq \epsilon_{\alpha} $, the total relaxation time 
at high temperature is
\begin{eqnarray}
\frac{1}{\tau(\epsilon)}
\simeq \frac{m^\ast D^2}{2\rho_{a} \hbar^3v^2_s} k_{B} T.
\end{eqnarray}
Therefore, in the EP regime, the IRT depends linearly on temperature.

Now we want to see how resistivity depends on 
temperature and $\alpha$ at low
temperature (BG regime) where $ \hbar \omega_{\bf q} \sim k_{B} T$.
In the BG regime, the IRT strongly decreased because the phonon 
population decreases exponentially for phonon absorption and the
sharp Fermi distribution prohibits phonon emission. 
To see the temperature dependence of the resistivity in the BG regime, 
it is convenient to calculate IRT averaged over energy,
as used for graphene \cite{graphene_t}, 
which is given by
\begin{equation}
\Big \la \frac{1}{\tau^{\lambda}} \Big \ra = 
\frac{\int d\epsilon 
D_{\lambda}(\epsilon ) \frac{1}{\tau^{\lambda}} 
[-\frac{df(\epsilon )}{d\epsilon }]}
{\int d\epsilon 
D_{\lambda} (\epsilon )[-\frac{df(\epsilon )}{d\epsilon }]}.
\end{equation}
Therefore, the resistivity of a 2DEG with RSOI can be calculated 
from the following equation:
\begin{equation}
\rho = \frac{m^\ast}{n_e e^2} \Big\la \frac{1}{\tau} \Big \ra,
\end{equation}
where $ \la 1/\tau \ra = \sum_{\lambda} \la 1/\tau^{\lambda} \ra $. 

When the temperature is very low we can make the following approximations:
(i)  the phonon energy is comparable with the thermal energy i.e 
$ k_{B} T \leq \hbar\omega_{\bf q} \ll \epsilon_{F}$ and
$f(\epsilon)[1-f(\epsilon \pm \hbar\omega_{\bf q})] 
\simeq \hbar\omega_{\bf q}( N_{\bf q} + 1/2 \pm 1/2) 
\delta(\epsilon-\epsilon_{F}).$ 
After taking these approximations, we obtain
\begin{eqnarray}
\Big \la\frac{1}{\tau^\pm} \Big \ra 
& \simeq &  \frac{2A D_0}{k_{B}T}
\Big(1\mp\sqrt{\frac{\epsilon_\alpha}{\epsilon_F+\epsilon_\alpha}}\Big) 
\nonumber\\ & \times & 
\int_{0}^{\pi} d\theta (1-\cos\theta)\vert C_{\bf q}\vert^2 \omega_{\bf q}
N_{\bf q}(N_{\bf q}+1).\nonumber\\
\end{eqnarray}
We can now convert the integration over $\theta$ into $q$ by using the relation
(based on small-angle scattering) 
$ q=2k_{F} \sin\frac{\theta}{2}$.
Substituting this and after a straightforward calculation we finally obtain
\begin{equation}
\Big \la\frac{1}{\tau^\pm} \Big \ra \simeq
\frac{D_0}{2}
\Big(1\mp\sqrt{\frac{\epsilon_\alpha}{\epsilon_{F}+\epsilon_\alpha}}\Big)
\frac{D^2}{\rho_{a} v_s}\frac{4! \zeta(4)}{(v_s\hbar)^4}
\frac{(k_{B}T)^4}{(k^{\pm}_F)^3},
\end{equation}
where $k_{F}^\pm = \sqrt{{k_{F}^0}^2 - k_\alpha^2 } \mp k_\alpha$, 
$k_\alpha=m^\ast\alpha/\hbar^2$ is the Rashba wave vector,
$k_{F}^0=\sqrt{2\pi n_e}$ is the Fermi wave vector of a 
2DEG without RSOI and $ \zeta(4) = \pi^4/(90)$.
In this temperature regime energy averaged IRT
is proportional to $T^4$ and hence $\rho \sim T^4$. 
The $T^4$ scaling law is also found in other perfect 2D system such as
graphene \cite{graphene_t}.
As we will see in the numerical calculations, the exponent
of the temperature dependence of the resistivity is reduced due
to SOI coupling.

\subsection{Numerical results}

In this section we determine the IRT, 
resistivity and their dependence on energy, temperature, SOI coupling 
constant etc. 
We solve Eq. (\ref{relax}) numerically.
For our numerical calculation,
we use $m^{\ast} = 0.067 m_0$ with $m_0$ is the free electron mass, 
$\alpha=\alpha_0=10^{-11}$ eV-m, $v_s=5.3\times10^{3} $ ms${}^{-1}$ and
the electron density $n_0 = 10^{15} $m$^{-2}$.

In Fig. 1 we have plotted IRT as a function of energy at a fixed 
temperature $T=1$ K for densities $n_e=n_0$ and 
$n_e=5n_0$. 
It could be seen from Fig. 1 that there is
a dip in the IRT. The dip occurs due to the sharpness of the Fermi distribution
function at low enough temperature. 
It is clear from Fig. 1 that the dip 
occurs exactly at $\epsilon=\epsilon_F^0$ when $\alpha = 0$. 
As $\alpha$ increases the position of the dip appearing at energies lower 
than $\epsilon_F^0$.
The dip occurs when $\epsilon=\mu$, where $\mu$ is the 
chemical potential which we evaluate numerically by solving Eq. (\ref{chem}). 
For the electron density $n_0$, the Fermi energy 
without RSOI is $\epsilon_F^0$=3.6 meV. 
When $\alpha=0$, $ \mu = \epsilon_F^0 = 3.6$ meV.
When $\alpha=3\alpha_0$, $\epsilon_F=2.8$ meV and 
$\mu = 0.8\epsilon_F^0$ so the dip 
occurs exactly at $\epsilon=0.8 \epsilon_F^0$. 
Similarly, for $\alpha=5\alpha_0$, $\epsilon_F=1.4$ meV and 
$\mu= 0.4\epsilon_F^0$ and consequently the dip occurs exactly at 
$\epsilon=0.4\epsilon_F^0$.
The IRT of the 2DEG system with RSOI is reduced compared to the absence of SOI.

\begin{figure}[t]
\begin{center}\leavevmode
\includegraphics[width=95mm]{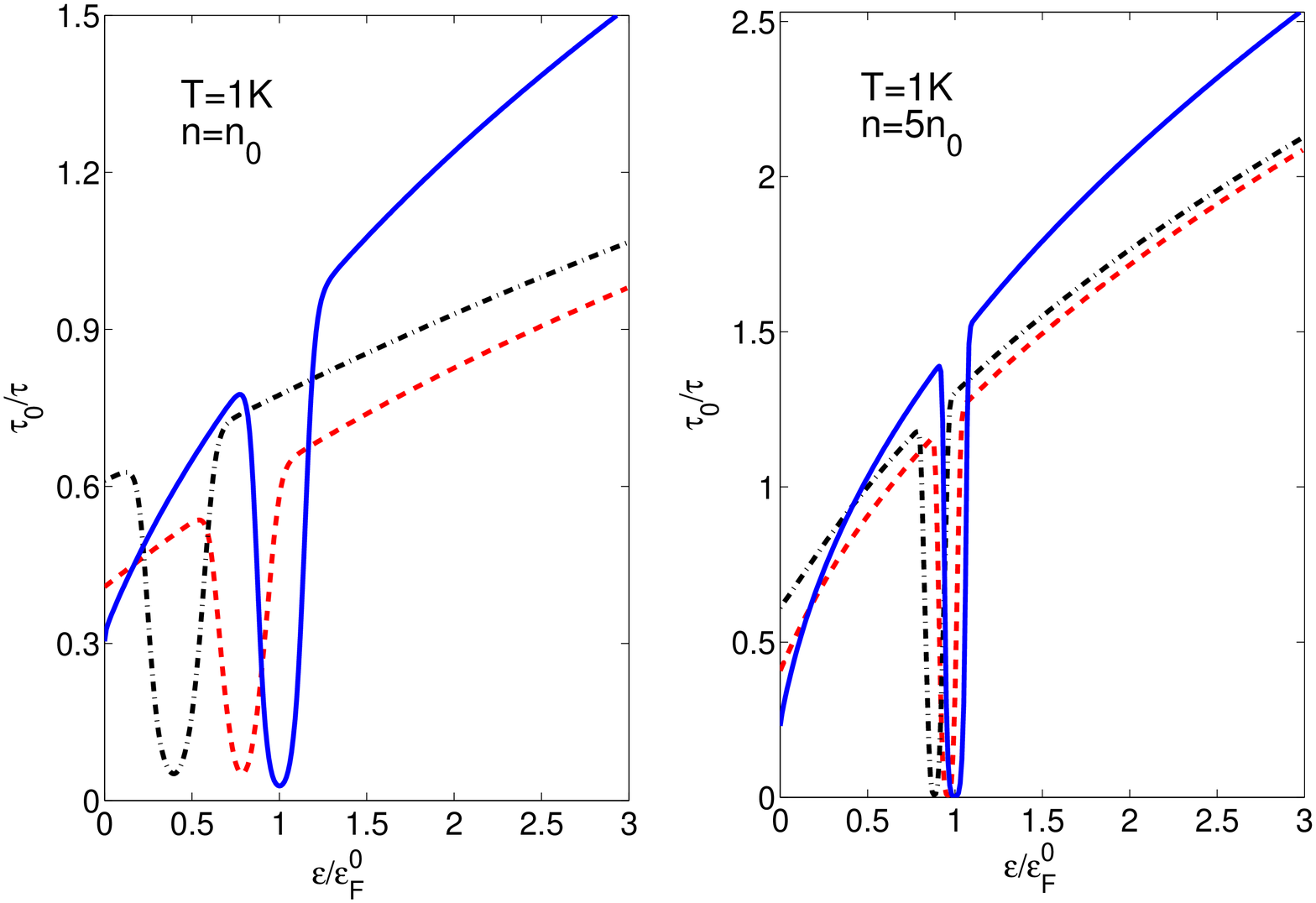}
\caption{(Color online) Plots of the IRT 
[in units of $1/\tau_0 = 2m^\ast D^2 \sqrt{2\pi n_0}/(\pi \hbar^2 \rho_{a} v_s)]$ 
versus energy for different values of $\alpha$. 
The solid, dashed and dot-dashed lines correspond to the $\alpha=0$, $\alpha = 3\alpha_0$
and $\alpha=5\alpha_0$, respectively. For better visualization, the solid lines in the 
left and the right panels have been reduced by a factor of 3 and 4, respectively.}
\label{Fig1}
\end{center}
\end{figure}

\begin{figure}[h!]
\begin{center}\leavevmode
\includegraphics[width=115mm]{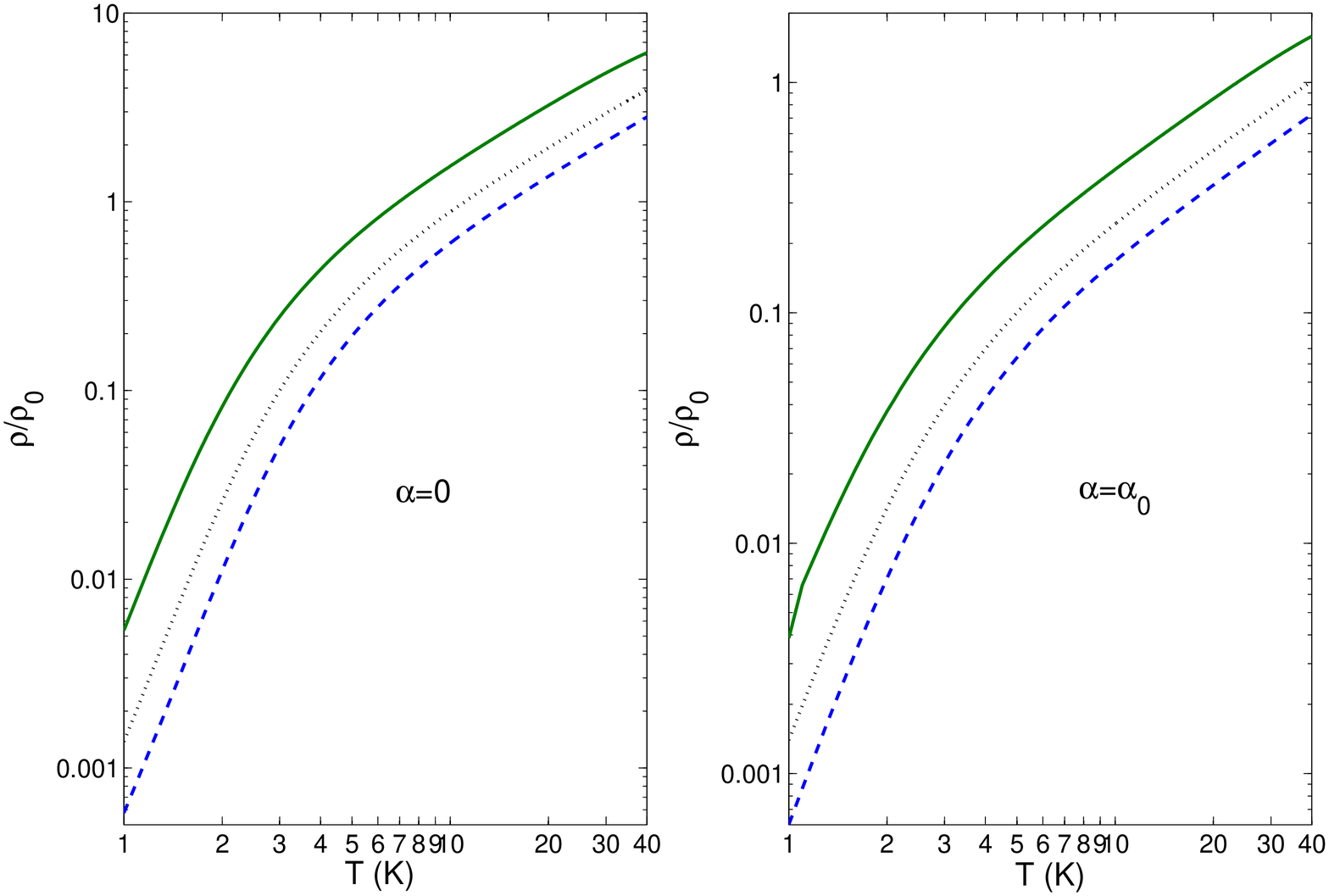}
\caption{(Color online) Plots of the resistivity 
[in units of $ \rho_0 = m^*/(n_0e^2 \tau_0)] $ of a 
perfect 2DEG versus $T$ for $\alpha=0$ and $\alpha=\alpha_0$ 
on a log-log scale. Here solid, dotted and dashed lines represent 
$n_e=3n_0 $, $n_e=5n_0$ 
and $n_e=7n_0$, respectively.}
\label{Fig2}
\end{center}
\end{figure}

\begin{figure}[ht]
\begin{center}\leavevmode
\includegraphics[width=120mm]{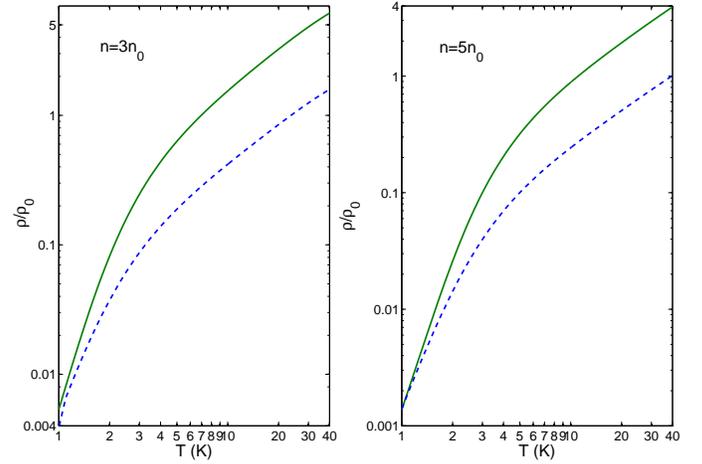}
\caption{(Color online) Plots of the resistivity with temperature
at a fixed density for different values of $\alpha$. Here,
solid and dashed lines correspond to the $\alpha =0 $ and
$\alpha = \alpha_0 $, respectively.}
\label{Fig3}
\end{center}
\end{figure}

\begin{figure}[ht]
\begin{center}\leavevmode
\includegraphics[width=120mm]{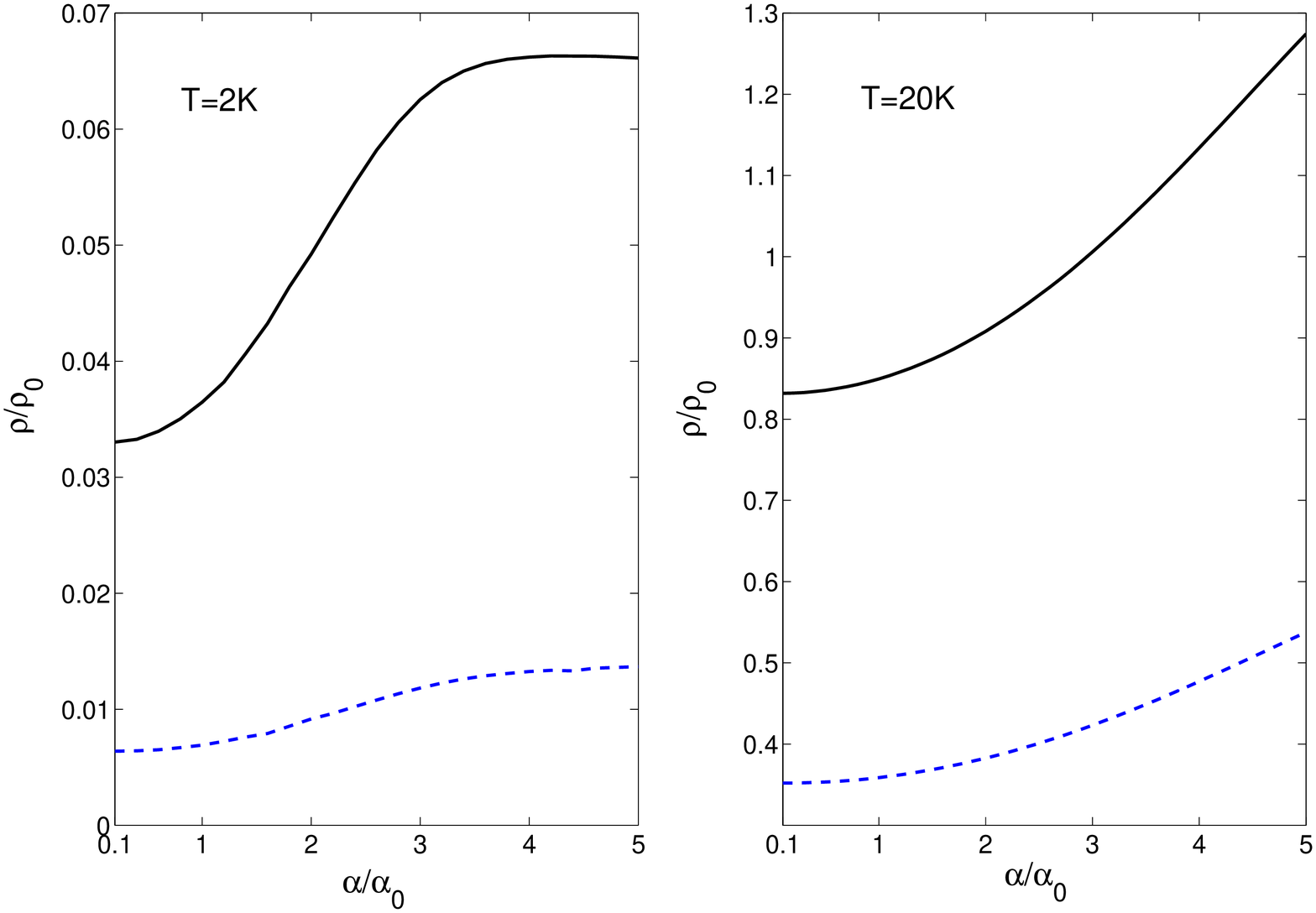}
\caption{(Color online) Plots of $\rho$ of a perfect 2DEG  
versus $\alpha$ for two fixed densities. 
Here solid and dashed lines correspond to 
$n_e = 3n_0$ and $n_e= 7n_0$, respectively.}
\label{Fig4}
\end{center}         
\end{figure}

The effective exponent ($\nu$) of the temperature dependence of 
the resistivity strongly depends on the density and the RSOI coupling constant. 
We estimate the effective exponent from the log-log plot of
the resistivity versus $T$ in Fig. 2 for 
$\alpha= 0$ and $\alpha = \alpha_0$
with different densities. 
In the BG regime ($T \sim 1-3$ K), we find
$\nu=3.763, 4.111 $ and $4.235$
for $n_e=3n_0, 5n_0$ and $7n_0$, respectively, when 
$\alpha = 0$. 
On the other hand, we find $\nu=2.829, 3.203$ and $3.433$ 
for $n_e=3n_0, 5n_0$ and $7n_0$, respectively, when 
$\alpha = \alpha_0$. The exponent $\nu$ is increasing with
electron density.

In the high temperature limit around $T = (18-40)$ K, we get 
$\nu=0.824, 0.998$ and $1.019$ for densities $n_e=3n_0, 5n_0$ and $7n_0$, 
respectively. The $\nu$ is also increasing 
with $n_e$ slowly and the equipartition result $\rho \sim T$ is recovered. 
The exponent $\nu$ does not change due to 
$\alpha$ in the high temperature limit.

In Fig. 3, we plot $\rho$ versus $T$ at a fixed density for different
values of $\alpha$. In the each panel of Fig. 3, we consider $\alpha=0$ 
and $\alpha=\alpha_0$ cases for densities $n_e=3n_0$ and $n_e=5n_0$. 
Figure 3 clearly shows that the slope of the curve for $\alpha=0$ 
is greater than that for $\alpha=\alpha_0$ case in the BG regime.
When $n_e=3n_0$, we estimate $\nu=3.763$ and $ \nu = 2.829 $ for 
$\alpha=0$ and $\alpha=\alpha_0$, respectively.
Similarly,  when $n_e=5n_0$ $\nu=4.111$ and $\nu=3.203$ for $\alpha = 0 $ 
and $\alpha=\alpha_0$, respectively, at very low temperature. 
The values of $\nu$ differ significantly between $\alpha=0$ and
finite $\alpha $ at very low temperature.

In Fig. 4, we plot resistivity of a perfect 2DEG  versus $\alpha$ 
for two different densities. In both the regimes, $\rho $ increases 
with $\alpha$. But the rate of increase of $\rho$ is high
for low electron density compared to that of the high density. 
In the BG regime, the resistivity is saturated at high $\alpha$.

\section{Electron-phonon scattering in a Quasi-2DEG}
In this section we consider three dimensional bulk phonon with 
wave vector ${\bf Q}=({\bf q},q_z)$ 
interacts with the two dimensional electron wave vector 
${\bf k}=(k_x,k_y)$. The component of $\bf Q$ 
in the $xy$ plane obeys the conservation of momentum 
i.e ${\bf k}-{\bf k^\prime}={\bf q}$ and the
$z$-component must be integrated out. 
In semiconductor heterostructure 
the electrons move in the $x$-$y$ plane in presence 
of a triangular potential in the
$z$ direction. It is also assumed that the lowest energy level
is occupied by the electrons. Generally the wave function can be written 
as $\psi({\bf r})=\psi(x,y)\zeta_0(z)$.
Here the variational wave function $\zeta_0(z)$ is given by, 
$\zeta_0(z)=\sqrt{b^3/2}ze^{-bz/2}$. 
The variational parameter \cite{ando} $b$ is given by 
$b=(48\pi m^\ast e^2/\varepsilon_0 \kappa_0 \hbar^2)^{1/3}
\Big(n_c+11n_e/32\Big)^{1/3}$, 
where $\kappa_0 = 12.9 $ is the static dielectric constant of GaAs, 
$\varepsilon_0$ is the free space permittivity and
$n_c$ is the depletion charge density in the channel. 
In this section we will also discuss the piezoelectric 
scattering along with the deformation potential scattering.
In this case the IRT can be written as
\begin{eqnarray} \label{quasi2d}
\frac{1}{\tau^{\lambda}(\epsilon)}&=& 
\frac{1}{(2\pi)^3}\frac{2\pi}{\hbar}\sum_{\lambda^\prime}
\int dk^\prime k^\prime\int d\theta(1-\cos\theta)\nonumber\\
&\times&
\int dq_z \vert I(q_z)\vert^2\vert 
C_{{\bf q},q_z}^{\lambda,\lambda^{\prime}} \vert^2\Big\{N_{Q}
\delta(\epsilon^\prime_{\lambda^\prime}-\epsilon_\lambda-\hbar\omega_{Q}) 
\nonumber\\ & + & (N_{Q}+1) 
\delta(\epsilon^\prime_{\lambda^\prime}-\epsilon_\lambda + \hbar\omega_{Q})\Big\}
\frac{1-f(\epsilon^\prime_{\lambda^\prime})}{1-f(\epsilon_\lambda)},
\end{eqnarray}
where the term $\vert I(q_z)\vert^2$, called the form factor, is responsible for
getting transition rate from three dimensional bulk phonon state to two dimension.
The exact form of this term for a triangular potential is given by
$\vert I(q_z)\vert^2= 
\vert \int dz \zeta_0^2(z)e^{iq_zz} \vert^2 = 
b^6/(b^2+q_z^2)^3$.

\subsection{ Deformation potential scattering}
The matrix element in this case is given by
\begin{equation}\label{dp_mat}
\vert C_{{\bf q},q_z}^{\lambda,\lambda^{\prime}} \vert^2=\frac{D^2\hbar Q}{2\rho_mv_s}
\frac{1 + \lambda \lambda^{\prime} \cos \theta}{2},
\end{equation}
where $\rho_m $ is the mass density. At high temperature, we have
$ N_{Q} \simeq N_{Q} + 1 \simeq  k_{B}T/(\hbar v_sQ)$.
Inserting this matrix element into Eq. (\ref{quasi2d}), we get 
\begin{eqnarray}
\frac{1}{\tau^\pm(\epsilon)}&\simeq&\frac{2m^\ast D^2}{\pi^2\hbar^3\rho_mv_s^2}
\Big(1\mp\sqrt{\frac{\epsilon_\alpha}{\epsilon+\epsilon_\alpha}}\Big)k_BT\nonumber\\
&\times & \int_0^1 dx x^2 \sqrt{1-x^2} 
\int_0^{\infty} dq_z\vert I(q_z)\vert^2,
\end{eqnarray}
with $x=q/2k$.
Performing the integrations over $x$ and 
$q_z$, we finally obtain the total IRT in the EP regime
for DP scattering as given by
\begin{equation}
\frac{1}{\tau(\epsilon)}=\frac{3}{32}\frac{m^\ast bD^2}{\rho_m\hbar^3v_{s}^2}k_BT.
\end{equation}

On the other hand, in the BG regime, we obtain the following expression 
for the energy averaged IRT (see appendix A1):
\begin{equation}
\Big\la\frac{1}{\tau^\pm}\Big\ra_{DP} \simeq \frac{D_0}{4}
\Big(1\mp\sqrt{\frac{\epsilon_\alpha}{\epsilon_F^0-\epsilon_\alpha}}\Big)
\frac{D^2}{\rho_mv_s}\frac{5!\zeta(5)}{(v_s\hbar)^5}\frac{(k_BT)^5}{(k^{\pm }_F)^3},
\end{equation} 
where $\zeta(5) = 1.037$.

\begin{figure}[ht]
\begin{center}\leavevmode
\includegraphics[width=115mm]{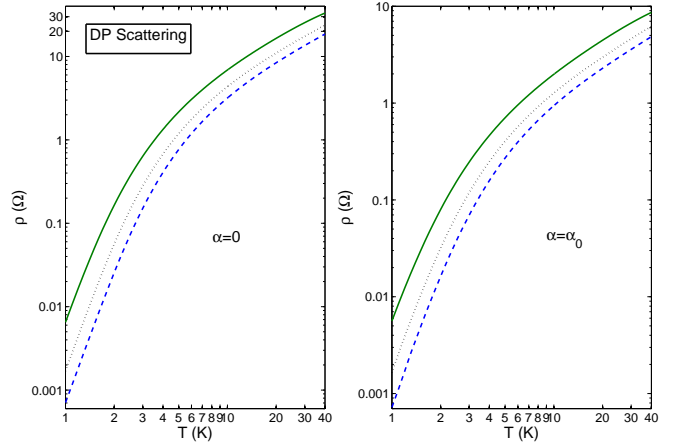}
\caption{(Color online) Plots of the resistivity 
of a quasi-2DEG due to DP scattering versus temperature
for different values of the density. Here, solid, dotted and dashed
lines represent $ n_e = 3n_0$, $n_e=5n_0$ and $ n_e = 7n_0$, respectively.}
\label{Fig5}
\end{center}
\end{figure}

\begin{figure}[ht]
\begin{center}\leavevmode
\includegraphics[width=115mm]{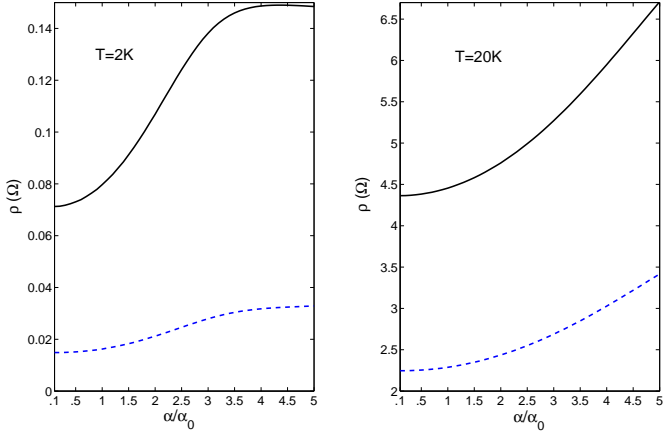}
\caption{(Color online) Plots of the resistivity of a quasi-2DEG 
due to DP scattering versus $\alpha$ for various densities. 
Here, solid and dashed lines correspond to $n_e=3n_0$ and
$n_e = 7n_0$, respectively.}  
\label{Fig6}
\end{center}         
\end{figure}

\subsection{Piezoelectric scattering}
Piezoelectricity is nothing but generation of polarization 
due to the application of a strain to a crystal without inversion
symmetry. Due to lattice vibration a potential can be generated in 
such crystals and electrons are scattered by this kind of potential. 
To calculate the IRT due to the PE scattering, we can use 
Eq. (\ref{quasi2d}). 
Following Ref. \cite{price1}, the matrix elements of the Rashba 
system are obtained as 
\begin{equation}\label{piezo_mat}
\vert C_{{\bf q},q_z,{l(t)}}^{PE, \lambda,\lambda^{\prime}}\vert^2=
\frac{(eh_{14})^2\hbar}{2\rho_mv_{s{l(t)}}} 
\frac{1 + \lambda \lambda^{\prime} \cos \theta}
{2\sqrt{q^2+q_z^2}}A_{l{(t)}}({\bf q},q_z),
\end{equation}
where $A_{l}({\bf q},q_z)=9q_z^2q^4/2(q_z^2+q^2)^3$ and 
$A_t({\bf q},q_z)=(8q_z^4q^2+q^6)/4(q_z^2+q^2)^3$.
In Eq. (\ref{piezo_mat}) the value PE tensor component $h_{14}$ is
$1.2\times10^9$ V/m and $v_{sl(t)}$ is the longitudinal (transverse) 
component of sound velocity.
By inserting this matrix element into Eq. (\ref{quasi2d}) and doing some 
straightforward calculations in the high temperature regime we get,
\begin{eqnarray}\label{long_IRT}
\frac{1}{\tau_l^\pm(\epsilon)}&\simeq&\frac{9}{8\pi^2}
\frac{m^\ast}{\hbar^3}\frac{(eh_{14})^2}{\rho_mv_{sl}^2}
\Big(1\mp\sqrt{\frac{\epsilon_\alpha}{\epsilon+\epsilon_\alpha}}\Big)
k_BT\nonumber\\
&\times&\int_0^\pi d\theta \sin^2\theta F_l(q)
\end{eqnarray}
and
\begin{eqnarray}\label{tran_IRT}
\frac{1}{\tau_t^\pm(\epsilon)}&\simeq&\frac{1}{16\pi^2}
\frac{m^\ast}{\hbar^3}\frac{(eh_{14})^2}{\rho_mv_{st}^2}
\Big(1\mp\sqrt{\frac{\epsilon_\alpha}{\epsilon+\epsilon_\alpha}}\Big)
k_BT\nonumber\\
&\times&\int_0^\pi d\theta \sin^2\theta F_t(q). 
\end{eqnarray}
Here $F_l(q)$ and $F_t(q)$ are respectively given by
\begin{eqnarray}
F_l(q)&=&\int dq_z\vert I(q_z)\vert^2\frac{q_z^2q^4}{(q_z^2+q^2)^4}\nonumber\\
&=&\frac{\pi}{16q}\frac{1+6\kappa+12\kappa^2+2\kappa^3}{(1+\kappa)^6}
\end{eqnarray}
and
\begin{eqnarray}
F_t(q) & = & \int dq_z \vert I(q_z)\vert^2 
\frac{8q_z^4q^2+q^6}{(q_z^2+q^2)^4} = \frac{\pi}{16q} \times \nonumber \\
& & \frac{13+78\kappa+72\kappa^2+82\kappa^3+36\kappa^4
+6\kappa^5}{(1+\kappa)^6},
\end{eqnarray}
where $\kappa=q/b$.
Now, we are assuming that the quasi-2DEG is very thin, i.e 
$\kappa \ll 1 $. So $F_l(q)$ and $F_t(q)$ can be
approximated as $F_l(q)\simeq \pi/16q$ and $F_t(q) \simeq 13\pi/16q$.

Substituting $F_l(q)$, $F_t(q)$ and  $q=2k \sin(\theta/2)$ in 
Eqs. (\ref{long_IRT}) and (\ref{tran_IRT}) and integrating 
over $\theta$, we obtain
\begin{equation}
\frac{1}{\tau_l^\pm(\epsilon)}\simeq\frac{3}{32}\frac{m^\ast}{\pi\hbar^3} 
\frac{(eh_{14})^2}{\rho_mv_{sl}^2}
\frac{1}{k^\pm}
\Big(1\mp\sqrt{\frac{\epsilon_\alpha}{\epsilon+\epsilon_\alpha}}\Big)k_BT
\end{equation}
and
\begin{equation}
\frac{1}{\tau_t^\pm(\epsilon)}\simeq\frac{13}{192}\frac{m^\ast}{\pi\hbar^3}
\frac{(eh_{14})^2}{\rho_mv_{st}^2}
\frac{1}{k^\pm}
\Big(1\mp\sqrt{\frac{\epsilon_\alpha}{\epsilon+\epsilon_\alpha}}\Big)k_BT.
\end{equation}

The total IRT for longitudinal and transverse cases are given as
\begin{equation}\label{long_PE}
\frac{1}{\tau_l (\epsilon)}\simeq\frac{3}{16}\frac{m^\ast}{\pi\hbar^3} 
\frac{(eh_{14})^2}{\rho_mv_{sl}^2}
\frac{k_BT}{k_F^0}
\sqrt{\frac{\epsilon_F^0}{\epsilon + \epsilon_\alpha}}
\end{equation}
and
\begin{equation}\label{trans_PE}
\frac{1}{\tau_t (\epsilon)}\simeq\frac{13}{96}\frac{m^\ast}{\pi\hbar^3}
\frac{(eh_{14})^2}{\rho_mv_{st}^2}
\frac{k_BT}{k_F^0}
\sqrt{\frac{\epsilon_F^0}{\epsilon + \epsilon_\alpha}}.
\end{equation}
The total IRT in the PE scattering case can be written as
$1/\tau^{PE}=1/\tau_l+2/\tau_t$.
In the high temperature regime, the IRT is proportional to $k_BT$
and inversely proportional to $\sqrt{\epsilon + \epsilon_{\alpha}}$.

In the low temperature regime we calculate IRT averaged over energy.
The detail calculations are given in appendix A2.
In this case, we have following expressions

\begin{eqnarray}\label{lon_t}
\Big\la\frac{1}{\tau_l^\pm}\Big\ra_{PE}&\simeq&\frac{45}{512}
\frac{m^\ast}{\pi\hbar}\frac{(eh_{14})^2}{\rho_m}
\frac{3!\zeta(3)}{(\hbar v_{sl})^4} \Big(\frac{k_BT}{k_F^\pm}\Big)^3 \nonumber\\
&\times&\Big(1\mp\sqrt{\frac{\epsilon_\alpha}{\epsilon_F^0  - \epsilon_\alpha}}\Big)
\end{eqnarray}
and 
\begin{eqnarray}\label{trn_t}
\Big\la\frac{1}{\tau_t^\pm}\Big\ra_{PE}&\simeq&\frac{59}{1024}\frac{m^\ast}{\pi\hbar}
\frac{(eh_{14})^2}{\rho_m}
\frac{3!\zeta(3)}{(\hbar v_{st})^4} \Big(\frac{k_BT}{k_F^\pm}\Big)^3 \nonumber\\
&\times&\Big(1\mp\sqrt{\frac{\epsilon_\alpha}{\epsilon_F^0 - \epsilon_\alpha}}\Big).
\end{eqnarray}
Here, $\zeta(3) = 1.202 $. 
% (Apéry's constant)
Equations (\ref{lon_t}) and (\ref{trn_t}) show that the energy averaged 
IRT due to PE scattering is proportional to  $T^3$ and 
hence the resistivity is also proportional to $T^3$.

\begin{figure}[ht]
\begin{center}\leavevmode
\includegraphics[width=115mm]{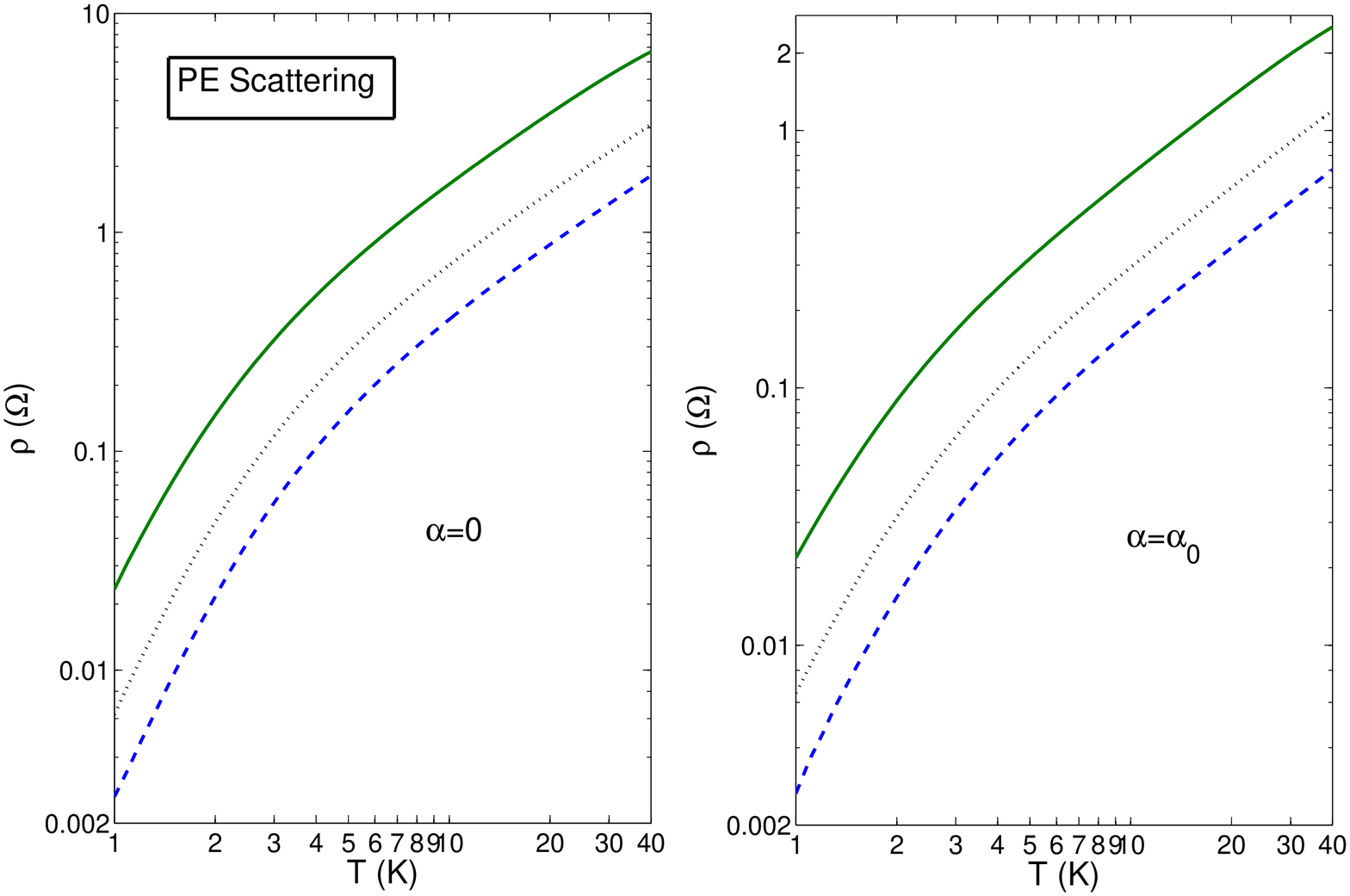}
\caption{(Color online) Plots of the resistivity due to PE scattering
versus temperature for different values of the density.
Here, solid, dotted and dashed lines represent
$n_e = 3n_0$, $n_e=5n_0$ and $ n_e = 7n_0$, respectively.}
\label{Fig7}
\end{center}
\end{figure}

\begin{figure}[]
\begin{center}\leavevmode
\includegraphics[width=115mm]{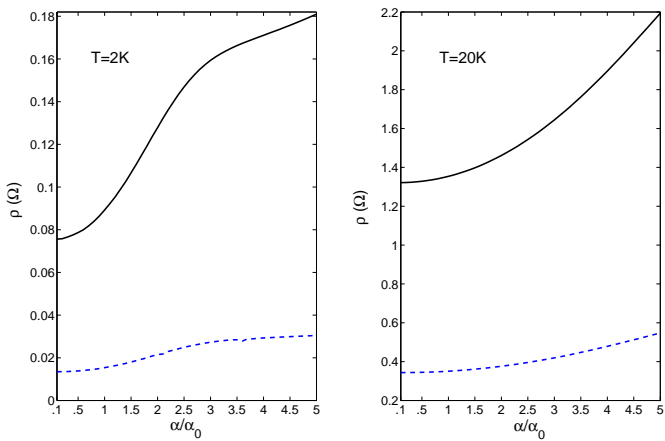}
\caption{(Color online) Plots of the resistivity of a
quasi-2DEG due to piezoelectric scattering
versus $\alpha$ for various densities.
Here, solid and dashed lines correspond to
$n_e=3n_0$ and $n_e=7n_0$, respectively.}
\label{Fig8}
\end{center}
\end{figure}

\subsection{Numerical results}
In this section we discuss resistivity due to both DP and 
PE scattering mechanisms. We solve Eq. (\ref{quasi2d}) numerically 
using the matrix elements for DP and PE scattering given in Eqs. 
(\ref{dp_mat}) and (\ref{piezo_mat}), respectively. 
For the numerical calculation we set 
$v_{sl} = 5.31 \times 10^3 $ ms${}^{-1}$, 
$v_{st}=3.04\times 10^3 $ ms${}^{-1}$, 
$n_c = 5 \times 10^{14}$ m${}^{-2}$, $\rho_m=5.12\times 10^3$ Kgm${}^{-3}$ and 
$D=12$ eV. 
The other parameters are the same as given in Sec. II.
The values of the numerical parameters considered in this section 
and also in Sec. II are appropriate for GaAs/AlGaAs heterostructure.
These material parameters are different for different kind
of heterostructures.

In Fig. 5, we show the temperature dependence 
of the resistivity of the quasi-2DEG 
due to the DP scattering for fixed values of $\alpha=0$
and $\alpha=\alpha_0$ with different densities.
In the BG regime with $\alpha=0$, 
the effective exponents of $T$ are $\nu= 4.459, 4.887$ and $5.085$
for densities $n_e=3n_0, 5n_0$ and $ 7n_0$, respectively.
Similarly, for $\alpha = \alpha_0$, 
we get the exponents as 
$\nu= 3.6499, 4.058$ and $4.359$ 
for densities $n_e=3n_0, 5n_0$ and $ 7n_0$, respectively. 
In this case,  $\nu$ is greater than the perfect 2DEG case 
discussed in the previous section.
This increase in $\nu$ is due to the finite thickness of quasi-2DEG
in the $z$ direction. 
In the EP regime, the $\nu$ is closed to one as we increase
$n_e$.

In Fig. 6, we plot the resistivity of a quasi-2DEG due to DP scattering 
as a function of $\alpha$ at fixed temperature and density.
Figure 6 shows that the $\rho$ 
increases rapidly with $\alpha$ in both the regimes but it is
faster for low electron density and gets saturated after 
certain value of $\alpha$ in the BG regime.

In Fig. 7, we plot the temperature dependence
of the resistivity of the quasi-2DEG
due to the PE scattering for fixed values of $\alpha= 0$
and $\alpha= \alpha_0$  with different densities.
When $\alpha =0$, the exponents are $\nu= 2.537, 2.830 $ and $3.002$ 
for densities $n_e=3n_0, 5n_0$ and $ 7n_0$, respectively, in the BG regime.
Similarly, the effective exponents are $\nu= 1.962, 2.208$ and $2.398$ for
densities $n_e=3n_0, 5n_0$ and $ 7n_0$, respectively, when 
$\alpha = \alpha_0$.
It clearly shows that the exponent $\nu$ is decreased due to the presence
of the RSOI.

For PE scattering case we also calculate
the resistivity  of a quasi-2DEG as a function of $\alpha$ by solving Eq. 
(\ref{quasi2d}) and plot the results in Fig. 8. 
Figure 8 shows that $\rho $ increases with $\alpha$ in both the regimes.
The other features are similar to the DP scattering mechanism.

Comparing Fig. 6 and Fig. 8, one could see that the resistivity
due to DP and PE potential are in the same order in
the BG regime but $\rho$ due to DP is dominating over PE in the 
EP regime.

In past most studies regarding phonon-limited transport phenomena
in heterostructures without spin-orbit interaction have been 
performed by focusing on mobility instead of resistivity. 
Our results for $\alpha=0$ case are consistent with those previous 
results.

\section{Summary}
In this work we have investigated the effect of 
the Rashba spin-orbit interaction on the momentum relaxation time due to
the electron-phonon scattering in a 2DEG. 
We have considered both perfect 2DEG and quasi-2DEG. 
We have also considered both the deformation potential and 
piezoelectric scattering 
mechanisms responsible for the electron-phonon interaction separately.
The temperature dependence of the resistivity has been
calculated in both equipartition and Bloch-Gruneisen regimes. 
We have found through approximate calculations that 
the resistivity of a perfect 2DEG is proportional to 
$T^4$ in the Bloch-Gruneisen regime for the deformation
potential scattering.
On the other hand, $\rho \sim T^5 $ for deformation potential 
scattering and $\rho \sim T^3$ for piezoelectric scattering in 
a quasi-2DEG. We have also recovered the linear temperature 
dependence of the resistivity in the equipartition regime of all the cases. 
Our numerical analysis showed that the effective exponent ($\nu$) of 
the temperature dependence of the resistivity {\it strongly} depends on 
the electron density $n_e$ and spin-orbit coupling constant $\alpha$. 
For the deformation potential and piezoelectric scattering, 
the values of the effective exponent of $T$ in the 
Bloch-Gruneisen regime for different values of $n_e$ and $\alpha$ are 
summarized in Table I.

\begin{table}[ht]
\centering
\begin{tabular}{|c|c|c|c|c|c|c|}\hline

\multicolumn{1}{|c|}{} &
\multicolumn{2}{c|}{ $\nu$ for 2DEG } &
\multicolumn{4}{c|}{$\nu $ for quasi-2DEG}\\ \cline{2-7}

\multicolumn{1}{|c|}{Density} &
\multicolumn{2}{c|}{DP} &
\multicolumn{2}{c|}{DP} &
\multicolumn{2}{c|}{PE} \\ \cline{2-7}

\multicolumn{1}{|c|}{$(n_e)$} &
\multicolumn{1}{|c|}{$\alpha=0$} &
\multicolumn{1}{|c|}{$\alpha=\alpha_0$} &
\multicolumn{1}{|c|}{$\alpha=0$} &
\multicolumn{1}{|c|}{$\alpha=\alpha_0$} &
\multicolumn{1}{|c|}{$\alpha=0$} &
\multicolumn{1}{|c|}{$\alpha=\alpha_0$} \\ \hline

\ \ \ 3$n_0$ & 3.763 & 2.829 & 4.459 & 3.649 & 2.537 & 1.962 \\
\ \ \ 5$n_0$ & 4.111 & 3.203 & 4.887 & 4.058 & 2.830 & 2.208 \\
\ \ \ 7$n_0$ & 4.235 & 3.433 & 5.085 & 4.359 & 3.002 & 2.398 \\ \hline

\end{tabular}
\caption{The effective exponent of the temperature dependence of 
the resistivity in the Bloch-Gruneisen regime for various values 
of $n_e$ and $\alpha$.}
\end{table}

There is a reduction in the exponent of temperature 
dependence of resistivity in Bloch-Gruneisen regime due to 
Rashba spin-orbit interaction at fixed electron density. 
We believe that the reduction in 
the exponent can be verified experimentally in near future.

The variation of $\rho$ with $\alpha$ has also been discussed 
for all the cases.It is found that the $\rho$ increases 
with $\alpha $ in both the regimes.
The rate of increase is faster in low electron density case.

\appendix
\section{}
In this appendix, we derive energy-averaged IRT due to DP and PE scattering 
in a quasi-2DEG in the BG regime.
At very low temperature, following the approximations used in section II(A)
and using Eq. (\ref{quasi2d}), we get
\begin{eqnarray}\label{q2dA}
\Big\la\frac{1}{\tau^\pm}\Big\ra & \simeq & 
\frac{m^\ast}{\pi^2\hbar^3}\Big(1\mp\sqrt{\frac{\epsilon_\alpha}
{\epsilon_F^0 - \epsilon_\alpha}}\Big) 
\frac{1}{k_BT}\int_{0}^{\pi} d\theta (1-\cos\theta)\nonumber\\
&\times&\int dq_z\vert I(q_z)\vert^2 
\vert C_{{\bf q},q_z}\vert^2\hbar\omega_{Q}N_{Q}(N_{Q}+1).
\end{eqnarray}
When $T$ is very low, the phonon wave vector $q$ is very very small 
compared to the Fermi wave vector i.e. $q \ll 2k_F$. 
For very thin quasi-2DEG,
$\vert I(q_z)\vert^2$ can be approximated as $\vert I(q_z)\vert^2\simeq 1$
since $ q_z \ll b$.
With all the approximations taken into account, 
Eq. (\ref{q2dA}) can be approximated further as 
\begin{eqnarray}\label{A2}
\Big\la\frac{1}{\tau^\pm}\Big\ra &\simeq& \frac{m^\ast}{2\pi^2\hbar^3}\frac{1}{k_F^{\pm3}}
\Big(1\mp\sqrt{\frac{\epsilon_\alpha}{\epsilon_F^0 - \epsilon_\alpha}}\Big)
\frac{1}{k_BT}\nonumber\\
&\times&\int dq dq_z q^2 \vert C_{{\bf q},q_z}\vert^2
\hbar\omega_{Q}N_{Q}(N_{Q}+1).
\end{eqnarray}

\subsection[]{Deformation Potential scattering}

Inserting the matrix element given in Eq. (\ref{dp_mat}) 
into Eq. (\ref{A2}), we get 
\begin{eqnarray}\label{A3}
\Big\la\frac{1}{\tau^\pm}\Big\ra_{DP} &\simeq & 
\frac{m^\ast D^2}{4\pi^2\hbar^3\rho_mv_s^2k_F^{\pm3}}
\Big(1\mp\sqrt{\frac{\epsilon_\alpha}{\epsilon_F^0 - \epsilon_\alpha}}\Big)
\frac{1}{k_BT}\nonumber\\
&\times&\int dq dq_z q^2(\hbar\omega_{Q})^2N_{Q}(N_{Q}+1).
\end{eqnarray}
Since the phonon dispersion relation is 
$\epsilon_p=\hbar v_s\sqrt{q^2+q_z^2}$,
we can make the following transformation:
$dqdq_z \rightarrow \epsilon_p d\epsilon_p 
d\phi/(\hbar v_s)^2$ with
$q= \epsilon_p\cos\phi/(\hbar v_s)$ and 
$q_z = \epsilon_p\sin\phi/(\hbar v_s)$.
With these transformations, Eq. (\ref{A2}) reduces to
\begin{eqnarray}
\Big\la\frac{1}{\tau^\pm}\Big\ra_{DP} & \simeq & 
\frac{m^\ast D^2}{4\pi^2\hbar^3\rho_mv_s^2k_F^{\pm3}}
\Big(1\mp\sqrt{\frac{\epsilon_\alpha}{\epsilon_F^0 - \epsilon_\alpha}}\Big)
\frac{1}{(\hbar v_s)^4}\nonumber\\
&\times&\frac{1}{k_BT}\int d\epsilon_p \epsilon_p^5N_Q(N_Q+1).
\end{eqnarray}

Using the result 
$\int d\epsilon_p\epsilon_p^nN_Q(N_Q+1)=n!\zeta(n)(k_BT)^{n+1}$ with
$\zeta(n)$ is the Riemann zeta function,
we finally obtain 
\begin{equation}
\Big\la\frac{1}{\tau^\pm}\Big\ra_{DP}\simeq\frac{D_0}{4}
\Big(1\mp\sqrt{\frac{\epsilon_\alpha}{\epsilon_F^0-\epsilon_\alpha}}\Big)
\frac{D^2}{\rho_mv_s}\frac{5!\zeta(5)}{(v_s\hbar)^5}\frac{(k_BT)^5}{(k^{\pm }_F)^3}.
\end{equation}

\subsection{Piezoelectric scattering}
Using the matrix elements for PE scattering given in Eq. (\ref{piezo_mat})
and Eq. (\ref{A2}), we obtain the energy-averaged IRT for longitudinal and 
transverse cases, respectively,
\begin{eqnarray}
\Big\la \frac{1}{\tau^\pm_l}\Big\ra_{PE} & \simeq & 
\frac{9}{8\pi^2}\frac{m^\ast (eh_{14})^2}
{\rho_m\hbar (\hbar v_{sl})^4 k_F^{\pm3}}
\Big(1\mp\sqrt{\frac{\epsilon_\alpha}{\epsilon_F^0-\epsilon_\alpha}}\Big)
\frac{1}{k_BT}\nonumber\\
& \times & 
\int d\epsilon_p\epsilon_p^3N_Q(N_Q+1)\int\cos^6\phi \sin^2\phi d\phi \nonumber
\end{eqnarray}
and
\begin{eqnarray}
\Big\la \frac{1}{\tau^\pm_t}\Big\ra_{PE} & \simeq & 
\frac{1}{8\pi^2}\frac{m^\ast (eh_{14})^2}
{\rho_m\hbar (\hbar v_{st})^4 k_F^{\pm3}}
\Big(1\mp\sqrt{\frac{\epsilon_\alpha}{\epsilon_F^0-\epsilon_\alpha}}\Big) 
\frac{1}{k_BT}\nonumber\\
&\times & 
\int(\cos^8\phi + 8\cos^4\phi \sin^4\phi)d\phi \nonumber \\
&\times & \int d\epsilon_p\epsilon_p^3N_Q(N_Q+1). 
\end{eqnarray}
After doing the integration over $\epsilon_p$ and $\phi$,
we finally obtain the following expressions
\begin{eqnarray}
\Big\la\frac{1}{\tau_l^\pm}\Big\ra_{PE}&\simeq&\frac{45}{512}
\frac{m^\ast}{\pi\hbar}\frac{(eh_{14})^2}{\rho_m}
\frac{3!\zeta(3)}{(\hbar v_{sl})^4} \Big(\frac{k_BT}{k_F^\pm}\Big)^3 \nonumber\\
&\times&\Big(1\mp\sqrt{\frac{\epsilon_\alpha}{\epsilon_F^0  - \epsilon_\alpha}}\Big)
\end{eqnarray}
and 
\begin{eqnarray}
\Big\la\frac{1}{\tau_t^\pm}\Big\ra_{PE}&\simeq&\frac{59}{1024}\frac{m^\ast}{\pi\hbar}
\frac{(eh_{14})^2}{\rho_m}
\frac{3!\zeta(3)}{(\hbar v_{st})^4} \Big(\frac{k_BT}{k_F^\pm}\Big)^3 \nonumber\\
&\times&\Big(1\mp\sqrt{\frac{\epsilon_\alpha}{\epsilon_F^0 - \epsilon_\alpha}}\Big).
\end{eqnarray}


\begin{thebibliography}{55} 

\bibitem{winkler}
R. Winkler, Spin-Orbit Coupling Effects in 
Two-Dimensional Electron and Hole Systems
(Springer Verlag-2003).

\bibitem{zutic}
F. Fabian, A. Matos-Abiague, C. Ertler, P. Stano, and
I. Zutic, 
Acta Physica Slovaca {\bf 57}, 565 (2007).

\bibitem{cahay}
S. Bandyopadhyay and M. Cahay, Introduction to
Spintronics (CRC press-2008).


\bibitem{datta}
S. Datta and B. Das,
Appl. Phys. Lett. {\bf 56}, 665 (1990).

\bibitem{fabian}
I. Zutic, J. Fabian, and S. Das Sarma,
Rev. Mod. Phys. {\bf 76}, 323 (2004).

\bibitem{wolf}
S. A. Wolf, D. D. Awschalom, R. A. Burhman, J. M. Daughton, S. von Molnar,
M. L. Roukes, A. Y. Chtchelkanova, and D. M. Treger, 
Science, {\bf 294}, 1488 (2001).

\bibitem{david}
D. D. Awschalom and M. E. Flatte, 
Nature physics {\bf 3}, 153 (2007).


\bibitem{rashba}
E. I. Rashba, Fiz. Tverd. Tela (Leningrad) {\bf 2}, 1224 (1960)
[Sov. Phys. Solid State {\bf 2}, 1109 (1960)];
Y. A. Bychkov and E. I. Rashba, J. Phys. C {\bf 17}, 580 (1984).


\bibitem{nitta}
J. Nitta, T. Akazaki, H. Takayanagi, and T. Enoki, 
Phys. Rev. Lett. {\bf 78}, 1335 (1997).

\bibitem{mats}
T. Matsuyama, R. Kursten, C. Meibner, and U. Merkt
Phys. Rev. B {\bf 61}, 15588 (2000).

\bibitem{dress}
G. Dresselhaus, Phys. Rev. {\bf 100}, 580 (1955).


\bibitem{cape}
E. Cappelluti, C. Grimaldi, and F. Marsiglio,
Phys. Rev. B {\bf 76}, 085334 (2007).

\bibitem{grima}
E. Cappelluti, C. Grimaldi, and F. Marsiglio,
Phys. Rev. Lett. {\bf 98}, 167002 (2007).

\bibitem{chap}
A. V. Chaplik and L. I. Magarill,
Phys. Rev. Lett. {\bf 96}, 126402 (2006).

\bibitem{mars}
C. Grimaldi, E. Cappelluti, and F. Marsiglio,
Phys. Rev. Lett. {\bf 97}, 066601 (2006).

\bibitem{chen}
L. Chen, Z. Ma, J. C. Cao, T.Y. Zhang, and C. Zhang,
Appl. Phys. Lett. {\bf 91}, 102115 (2007).

\bibitem{zhang}
Z. Li, Z. S. Ma, A. R. Wright, and C. Zhang,
Appl. Phys. Lett. {\bf 90}, 112103 (2007).

\bibitem{pagano}
P. M. Krstajic, M. Pagano, and P. Vasilopoulos,
Physica E {\bf 43}, 893 (2011).

\bibitem{knight}
D. L. Rode and S. Knight,
Phys. Rev. B {\bf 3}, 2534 (1971).

\bibitem{jiang}
C. Jiang, D. C. Tsui, and G. Weimann,
Appl. Phys. Lett. {\bf 53}, 1533 (1988).

\bibitem{gold}
A. Gold, 
Appl. Phys. Lett. {\bf 54}, 2100 (1989).

\bibitem{dolgo}
A. Gold and T. Dolgopolov,
J. Phys.: Condens. Matter {\bf 14}, 7091 (2002).

\bibitem{loss}
O. Chalaev and D. Loss,
Phys. Rev. B {\bf 80}, 035305 (2009).

\bibitem{stormer}
H. L. Stormer, L. N. Pfeiffer, K. W. Baldwin, and K. W. West,
Phys. Rev. B {\bf 41}, 1278 (1990).


\bibitem{price1}
P. J. Price,
Ann. Physics (N.Y) {\bf 133} 217 (1981).

\bibitem{price2}
P. J. Price,
J. Vac. Sci. Technol. {\bf 19}, 599 (1981).

\bibitem{price3}
P. J. Price,
Surf. Sci. {\bf 113}, 199 (1982).

\bibitem{ridley}
B. K. Ridley, 
J. Phys. C: Solid State Phys. {\bf 15}, 5898 (1982).

\bibitem{price4}
P. J. Price,
Surf. Sci. {\bf 143}, 145 (1984).

\bibitem{price5}
P. J. Price,
Solid State Commun. {\bf 51}, 607 (1984).


\bibitem{dassarma}
T. Kawamura and S. Das Sarma,
Phys. Rev. B {45}, 3612 (1992).

\bibitem{mos2}
K. Kaasbjerg, Antti-Pekka Jauho, and K. S. Thygesen,
arXiv: 1206:2003v1.

\bibitem{graphene_t}
E. H. Hwang and S. Das Sarma,
Phys. Rev. B {\bf 77}, 115449 (2008).

\bibitem{castro}
E. V. Castro, H. Ochoa, M. I. Katsnelson, R. V. Gorbachev,
D. C. Elias, K. S. Novoselov, A. K. Geim, and F. Guinea,
Phys. Rev. Lett. {\bf 105}, 266601 (2010).


\bibitem{kris}
K. Kaasbjerg, K. S. Thygesen, and K. W. Jacobsen,
Phys. Rev. B {\bf 85}, 165440 (2012).


\bibitem{graphene_exp}
D. K. Efetov and P. Kim,
Phys. Rev. Lett. {\bf 105}, 256805 (2010).

\bibitem{gao}
F. Gao, D. J. D. Beaven, J. Fulcher, C. H. Yang, Z. Zeng, W. Xu, and C. Zhang,
Physica E {\bf 40}, 1454 (2008).

\bibitem{ando}
T. Ando, Alan B. Fowler, and F. Stern,
Rev. Mod. Phys. {\bf 54}, 437 (1982).



\end{thebibliography}
\end{document}